\documentclass[twocolumn,prd,floatfix,superscriptaddress,preprintnumbers]{revtex4}

\usepackage{times} %

\usepackage{graphicx} %
\usepackage{color}%
\usepackage{caption,subfig}

\usepackage{amsmath,amssymb,slashed} %

\IfFileExists{srcltx.sty}{\usepackage[active]{srcltx}}

\usepackage{bm,paralist,xspace} %

\newcommand{\cc}{{\textsc{cc}}} %
\newcommand{\nc}{{\textsc{nc}}} %
\newcommand{\tr}{{\mathrm{tr}}} 
\newcommand{\mev}{\:\mathrm{MeV}} 
\newcommand{\gev}{\:\mathrm{GeV}} 
\newcommand{\parfrac}[2]{\left(\frac{#1}{#2}\right)}

\newcommand{\Ltot}{L_{\rm tot}} %
\newcommand{\cs}{\textsc{cs}} %
\newcommand{\sw}{\sin^2\theta_W} %

\begin{document}

\title{Long-range magnetic fields in the ground state of the Standard Model
  plasma}

\author{Alexey~Boyarsky}%

\affiliation{Instituut-Lorentz for Theoretical Physics, Universiteit Leiden,
  Niels Bohrweg 2, Leiden, The Netherlands} 

\affiliation{Ecole Polytechnique F\'ed\'erale de Lausanne, FSB/ITP/LPPC, BSP
  720, CH-1015, Lausanne, Switzerland}

\affiliation{Bogolyubov Institute of Theoretical Physics, Kyiv, Ukraine} %

\author{Oleg Ruchayskiy}%
\affiliation{CERN Physics Department, Theory Division, CH-1211 Geneva 23,
  Switzerland}

\author{Mikhail Shaposhnikov}
\affiliation{Ecole Polytechnique F\'ed\'erale de Lausanne, FSB/ITP/LPPC, BSP
  720, CH-1015, Lausanne, Switzerland}

\date{\today}

\begin{abstract}
  In thermal equilibrium the ground state of the plasma of Standard Model
  particles is determined by temperature and exactly conserved combinations of
  baryon and lepton numbers. We show that at non-zero values of the global
  charges a translation invariant and homogeneous state of the plasma
  becomes unstable and the system transits into a new state, containing a
  large-scale magnetic field.  The origin of this effect is the
  parity-breaking character of weak interactions and chiral anomaly.  This
  situation can occur in the early Universe and may play an important role in
  its subsequent evolution.
\end{abstract}
\maketitle


It is generally believed that the ground state of the Standard Model at high
temperatures is homogeneous and isotropic.  This assumption underlies the
description of all the important processes in the early Universe:
baryogenesis, cosmological phase transitions, primordial nucleosynthesis,
etc.\footnote{The primordial density fluctuations are considered as tiny at
  these epochs and take place only in the ``dark'' sector, decoupled from the
  equilibrium Standard Model plasma.}
In this work we demonstrate, however, that at finite density of lepton or
baryon numbers due to parity-violating nature of the weak interactions this
homogeneous ``ground state'' becomes unstable by developing a long-range
\emph{magnetic field}.  The transition to the ``true'' ground state may depend
on the details of the non-equilibrium dynamics, when various violent
dissipative processes (e.g.  turbulence, radiation emission, finite
conductivity of plasma) play an important role.

What are the conditions for the translational invariance to be spontaneously
broken by a long-range field? It is sufficient for the free energy of the
gauge fields to contain an interaction term that dominates over the kinetic
energy and can be both positive and negative. An example is provided by a
\emph{Chern-Simons} term $I_\cs \propto A \partial A$, that has less
derivatives that the kinetic term $(\partial A)^2$ and therefore can dominate
over it at large scales.  The presence of the Chern-Simons term in the Maxwell
equations is known to lead to an instability and generation of magnetic
fields.

At zero temperatures and densities the Chern-Simons term for electromagnetic
fields is prohibited as a consequence of gauge invariance and Lorentz symmetry
(\emph{Furry theorem}~\cite{Furry:1937zz}). At finite temperatures and
densities the plasma creates a preferred reference frame and the 4-dimensional
Lorentz invariance is broken down to 3-dimensional one.  As a result the free
energy of static gauge fields is
\begin{equation}
  \label{eq:8}
  \mathcal{F}[A] =  \int d^3p\, A_i(\vec p) \Pi_{ij}(p) A_j(-\vec p) +\mathcal{O}(A^3)
\end{equation}
with the \emph{polarization operator}
\begin{equation}
  \label{eq:63}
  \Pi_{ij}(\vec p) = (p^2\delta_{ij} - p_i p_j)\Pi_1(p^2) 
  + i \epsilon_{ijk}p^k \Pi_2(p^2)\;,
\end{equation}
where $i,j,k=1,2,3$ are spacial indices; $p^2 = |\vec p|^2$; $\epsilon_{ijk}$
is the antisymmetric tensor.  Eq.~(\ref{eq:63}) is the most general form of
$\Pi_{ij}$ satisfying the gauge-invariance transversality condition $p_i
\Pi_{ij} = 0$. In the long wavelength limit $p^2 \to 0$ a non-zero $\Pi_2(0)$
means that the Chern-Simons term $\Pi_2(0) \vec A \cdot \vec \nabla\times \vec
A$ appears in~(\ref{eq:8}). The $3\times3$ matrix~(\ref{eq:63}) has then a
negative eigenvalue for sufficiently small momenta $p <
|\Pi_2(p^2)/\Pi_1(p^2)|$ and the corresponding eigenmode grows larger and
larger (until the higher order in $A$ terms would stabilize it). In the above
consideration it is important that the gauge field is Abelian. Unlike the
Yang-Mills fields~\cite{Linde:1980ts} the magnetic component of the photon
field does not get screened in plasma~\cite{Fradkin:65} (i.e. $\Pi_1(0)$
remains finite) and therefore the instability does not require large
$\Pi_2(0)$.

In this work we demonstrate that in the Standard Model plasma in the Higgs
phase an \emph{equilibrium} value of $\Pi_2(0)$ for electromagnetic fields is
non-zero and proportional to the values of the global charges: \emph{baryon}
($B$) and \emph{flavor lepton} numbers $L_\alpha$ (index $\alpha$ runs over
flavours).  Unlike the previous
works~\cite{Redlich:1984md,Rubakov:86b,Rubakov:86,Deryagin:1986kx,Joyce:97}
(see discussion below) it is important that even if the anomalous charge $B+L$
is absent, $\Pi_2(0)$ remains non-zero and magnetic fields develop
\footnote{Moreover, at temperatures below sphaleron freeze-out that we
  consider here, the rate of the violation of $B+L$ is exponentially
  suppressed~\cite{Kuzmin:85}, therefore we can treat it as a conserved charge
  along with $L_\alpha - B/3$.}.

\smallskip

\begin{figure*}
  \begin{tabular}[c]{cccc}
    \subfloat[]
    {\label{fig:Redlich}%
      \includegraphics[width=.23\textwidth]{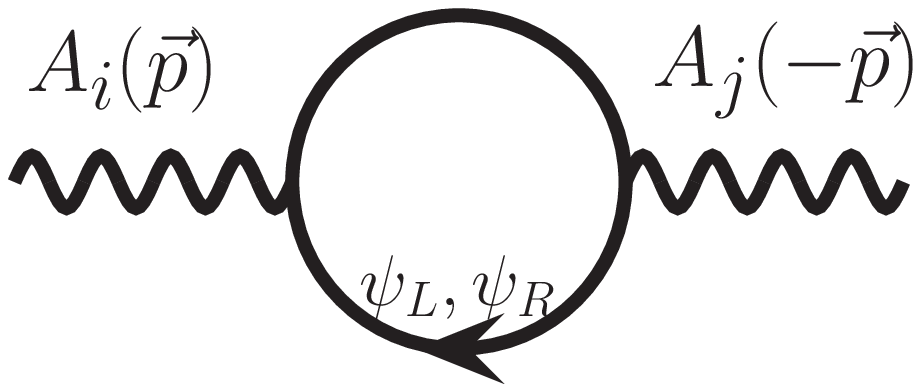}}
    &
    \subfloat[]
    {\label{fig:bubble}%
      \includegraphics[width=.23\textwidth]{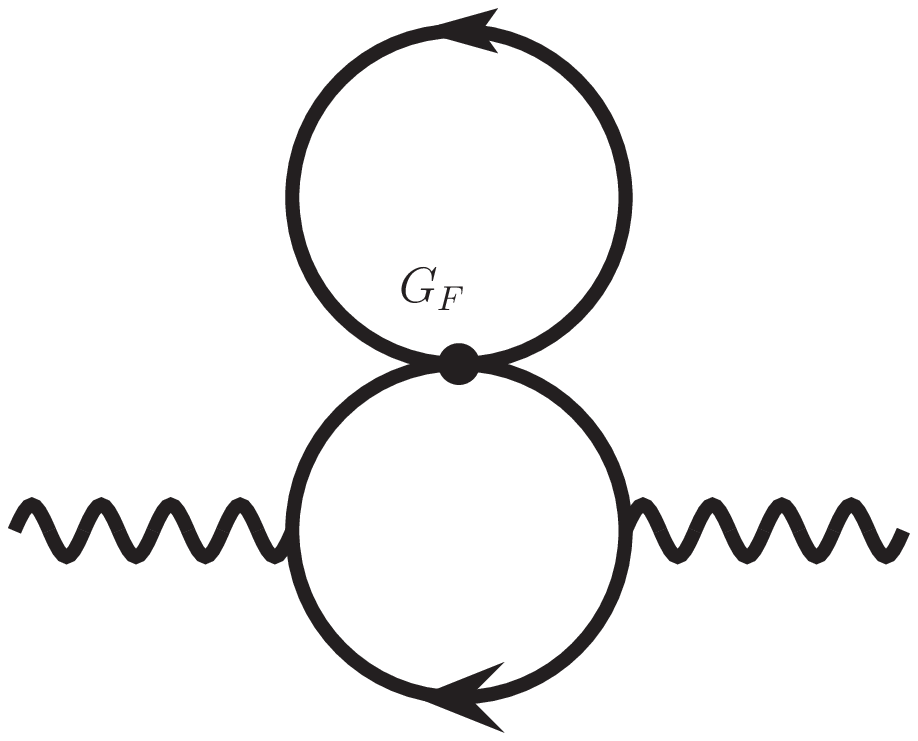}}
    &
    \subfloat[]
    {\label{fig:vertex}%
      \includegraphics[width=.23\textwidth]{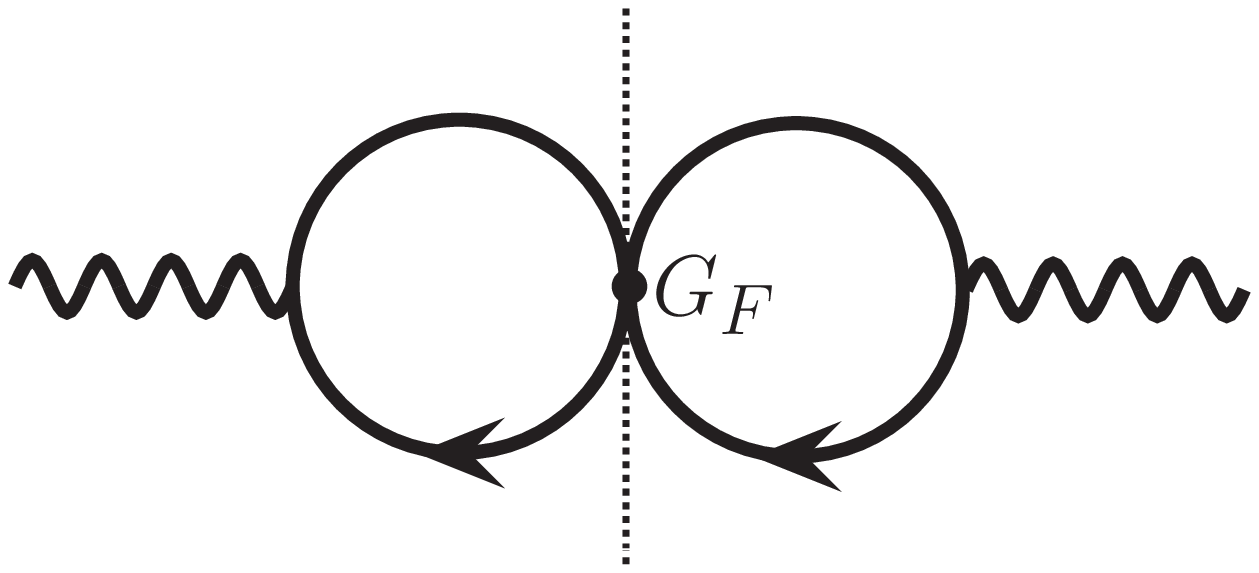}}
    &
    \subfloat[]
    {\label{fig:mu}\includegraphics[width=.23\textwidth]{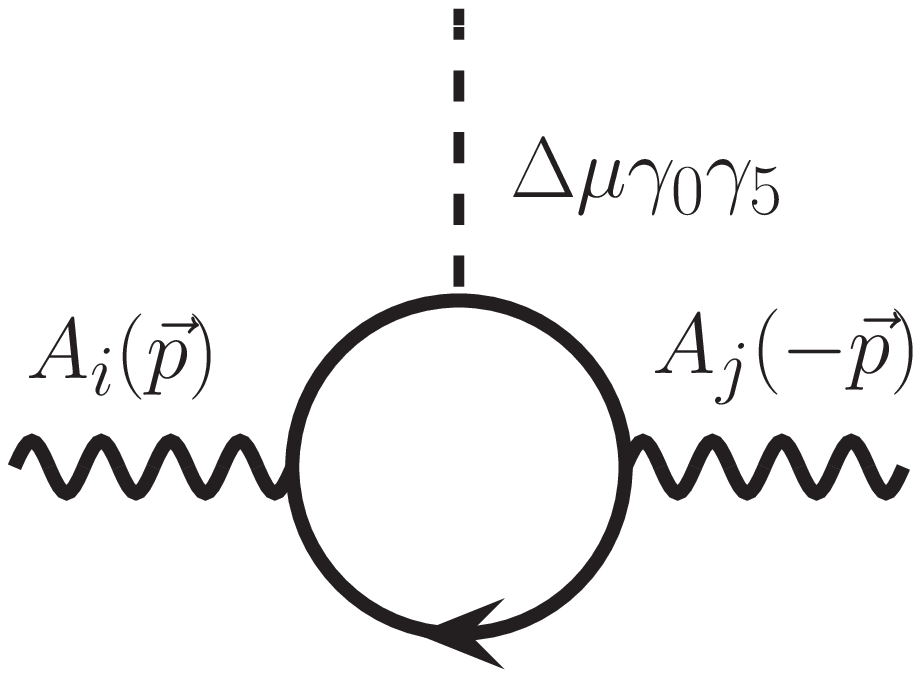}}  
  \end{tabular}
  \caption{Polarization operator \protect\subref{fig:Redlich}, one-loop weak
    corrections to it (\protect\subref{fig:bubble} and
    \protect\subref{fig:vertex}) and its expansion in $\Delta \mu/T$
    \protect\subref{fig:mu}.\label{fig:polarization}}
\end{figure*}

\noindent\textbf{Chern-Simons term and axial anomaly.}
The origin of the $\Pi_2$ term has its roots in the axial anomaly (see
e.g.~\cite{Redlich:1984md,Tsokos:85,Niemi:85,Niemi:86,Rubakov:86}). Indeed,
the non-conservation of the axial current at finite densities of left or right
fermions $n_L,n_R$ means that one can convert fermions into gauge field
configurations with a non-trivial \emph{Chern-Simons number} $N_\cs \equiv\int
d^3 x\, A\cdot B$ (where $B = \nabla \times A$ is a magnetic field):
\begin{equation}
  \label{eq:4}
  \frac{d (n_L - n_R)}{dt} = \frac{e^2}{2\pi^2}  \int d^3 x\, E \cdot B  =
  \frac{\alpha}{\pi} \frac{dN_\cs}{dt} 
\end{equation} 
(here $E = -\dot A$ is an electric field and $\alpha = \frac{e^2}{4\pi}$ is
the fine-structure constant).  Let us consider the simplest example of left
and right fermions at zero temperature with different Fermi energies (chemical
potentials) $\mu_L\neq \mu_R$. Infinitesimal change of the gauge field $\delta
A$ will destroy (create) $\delta n_{L,R} = \pm \frac{\alpha}{2\pi}\delta
N_\cs$ of real fermions around the Fermi level. The total energy of the system
will change by $\delta \mathcal{F} = (\mu_L - \mu_R) \frac{\alpha}{2\pi}\delta
N_\cs$~\cite{Rubakov:86}, which leads to the parity-odd Chern-Simons term in
the free energy: $\mathcal{F}[A] = \frac{\alpha(\mu_L - \mu_R)}{2\pi}\int
d^3x\, A\cdot B$.

This remains true in any vector-like gauge theory at finite
temperature/density where there is a difference of chemical potentials of left
and right-chiral charged
particles~\cite{Vilenkin:80a,Redlich:1984md,Tsokos:85}
(also~\cite{Alekseev:98a,Frohlich:2000en}).  Indeed, to calculate the
polarization operator~(\ref{eq:63}) we need to analyze one-loop contribution
from charged fermions described by the diagram~\ref{fig:Redlich}.  If the left
and right fermions have different chemical potentials such that
\begin{equation}
  \label{eq:5}
  G_{L,R} = \frac1{\gamma_0(i\omega_n + \mu_{L,R}) + \gamma \cdot p}P_{\text{L,R}}
\end{equation}
(where $\omega_n = \pi (2n+1) T$, $n\in\mathbb{Z}$ are the Matsubara
frequencies and $P_{\text{L,R}} = \frac12(1\pm \gamma_5)$ are chiral
projectors) their contributions to this diagram are different.  Assuming first
that $\Delta\mu\ll T $, let us consider linear in $\Delta\mu/T$, correction to
the polarization operator (for this one should differentiate the fermions
Green’s function~(\ref{eq:5}) with respect to $\mu$). This correction is
described by the diagram~\ref{fig:mu}, with $\Delta\mu$ playing the role of a
third external field.  The diagram~\ref{fig:Redlich} thus turns into the
famous triangular graph for the axial
anomaly~\cite{Adler:1969gk,Bell:1969ts,Treiman:85}, with the third vertex
containing ``axial vector field'' $X_\beta = \delta_{\beta0}\Delta
\mu\gamma_0\gamma_5$. The resulting term in the effective action,
$\propto\epsilon_{\alpha\beta\mu\nu} X_\alpha A_\beta \partial_\mu A_\nu$,
again reduces to the Chern-Simons term with $\Pi_2(0) =
\frac{\alpha}{2\pi}\Delta \mu$. This expression for $\Pi_2(0)$ is actually
exact in $\Delta \mu$ and $T$~\cite{Redlich:1984md}.

Similar logic applies to the non-Abelian gauge
fields~\cite{Tsokos:85,Niemi:85,Niemi:86,Rubakov:86b,Deryagin:1986kx}. In the
Standard Model with its chiral weak charges of fermions the coefficient in
front of the SU(2) Chern-Simons term can be expressed in terms of $\mu_{B +
  L}$ ($B$ being baryon and $L$ lepton
numbers)~\cite{Deryagin:1986kx,Gynther:03}.  However, in this case a
homogeneous state becomes unstable only at large values of chemical potential,
exceeding the mass of weak bosons. Even at high temperatures in the symmetric
phase the ``magnetic screening'' effect~\cite{Linde:1980ts} requires
$\Delta\mu \gtrsim T$ to overcome the ``magnetic mass'' $m_{magn}\sim \alpha_W
T$.  Moreover, anomalous non-conservation of $B + L$ current drives the
coefficient of corresponding Chern-Simons term to zero~\cite{Kuzmin:85} and
the standing wave-like configurations of the gauge fields are actually
metastable~(see discussion in \cite{Deryagin:1986kx}).

An example of the situation when the non-zero coefficient in front of the
Chern-Simons is realized in the Standard Model at high temperatures, $T\gg
m_f$, when the smallness of the electron's Yukawa coupling makes the number of
\emph{right electrons} conserved at classical level. If the initial conditions
have non-zero $\mu_{e_R}$, the Chern-Simons term for the U(1) hyper-field is
then generated (with $\Pi_2(0) \propto \mu_{e_R}$) and the generation of long
wave-length magnetic fields occurs~\cite{Joyce:97}, until the
chirality-flipping reactions, suppressed as $(m_e/T)^2$ do not destroy the
$\mu_{e_R}$.  In the early Universe where such a situation can be realized,
the rate of these reactions becomes comparable to the Hubble expansion rate at
high temperatures $T\sim 80$~TeV.  Magnetic fields, generated in such a way
are rather short-wavelength (much smaller than the horizon size at that epoch)
and are probably erased during the subsequent evolution due to plasma
dissipative processes.  At lower temperatures all the chirality-flipping
reactions are in thermal equilibrium and naively the chemical potentials of
all left and right-chiral particles are equal.  However, it was shown
in~\cite{Boyarsky:11a} that if strong helical magnetic fields are initially
present in the plasma, then the relaxation rate both for $\Delta \mu$ for
electrons and for helical fields significantly increases and they both can
survive down to $T\sim 10\mev$. The ground state that the system eventually
reaches contains neither fields nor $\Delta\mu$.

We demonstrate below that although these considerations are true for
electrodynamics, in the Standard Model where fermions are also involved in
parity-violating weak interactions, the difference of chemical potentials of
\emph{all} left- and right-chiral fermions is actually present (with all
chirality-flipping reactions taken into account) and leads to the generation
of magnetic fields.

In this paper we analyze the simplest situation when this effect is present:
the case $T\ll m_W$ (mass of the $W$-boson) when weak interactions can be
described by the Fermi theory:
\begin{equation}
  \label{eq:60}
  \mathcal{L}_\text{F} = \frac{4 G_F}{\sqrt 2}[(J_\mu^\nc)^2+2
  (J_\mu^\cc)^2]\;.
\end{equation}
The full Hamiltonian of the theory $\mathcal{H} = \mathcal{H}_0 +
\mathcal{H}_\text{F} + \mathcal{H}_{EM}$ has a free part for fermions and
photons, $\mathcal{H}_0$, and terms describing electromagnetic
($\mathcal{H}_\text{EM}$) and Fermi ($\mathcal{H}_\text{F}$) interactions
\footnote{ Here $G_F = 1.166\times 10^{-5}\gev^{-2}$ is the Fermi coupling
  constant and $ J^{\cc,\pm}_\mu$, $ J^\nc_\mu$ are \emph{charged} and
  \emph{neutral} currents (see definitions e.g.  in~\cite[Chapter
  20]{Peskin-Schroeder}).}.

\begin{figure}[!t]
\includegraphics[width=.4\textwidth]{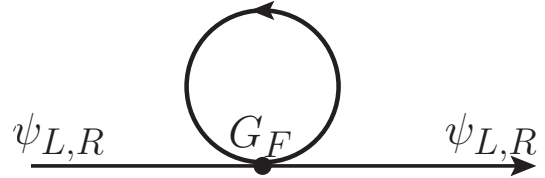} 
\caption{Fermi corrections to the self-energy of the fermion. The loop gives
  non-zero contribution only at finite lepton and baryon number density.}
\label{fig:self-energy-graph}
\end{figure}

\medskip

\noindent\textbf{Dispersion relation of fermions and chemical potentials.}
To describe the equilibrium plasma at $T\ll m_W$ we introduced the density
matrix, $\hat \varrho = \mathcal{Z}^{-1}\exp\Bigl(-\beta \bigl(\mathcal{H} -
\sum\limits_{\alpha}\lambda_\alpha L_\alpha - \lambda_Q Q -\lambda_B
B\bigr)\Bigr)$.  Five global charges commute with the Hamiltonian
$\mathcal{H}$: three $ L_\alpha$, $ B$ and $Q$
($\lambda_\alpha,\lambda_Q,\lambda_B$ are the corresponding Lagrange
multipliers); and the partition function $\mathcal{Z}$ ensures that
$\tr(\hat\varrho) = 1$.

To find the distribution functions of left and right-chiral particles we
compute the correlators $\langle \bar\psi P_{\text{L,R}}\psi\rangle =
\tr(\hat\varrho \bar\psi P_{\text{L,R}}\psi)$.  We expand the density matrix
in interactions to get
\begin{equation}
  \label{eq:2}
  \hat\varrho \approx \hat\varrho_0 \left(1 -\beta
    \mathcal{H}_\text{F} -\beta \mathcal{H}_\text{EM} - \frac{\beta^2}2
    \mathcal{H}_\text{EM} \mathcal{H}_\text{F} +\dots\right)\;.
\end{equation}
At zeroth order in interactions one gets $\langle \bar\psi P_{L}\psi\rangle_0
= \langle \bar\psi P_{R}\psi\rangle_0 = \frac12 \langle \bar\psi
\psi\rangle_0$.  The propagators of left and right particles (up to
corrections of the order $m/T\ll 1$) have the form~(\ref{eq:5}) with $\mu_L =
\mu_R$. This conclusion remains true if we take into account parity-preserving
electromagnetic interactions.

Taking into account chiral Fermi interactions, $\hat\varrho_F \approx
\hat\varrho_0(1-\beta \mathcal{H}_\text{F})$ one finds that $ \langle \bar\psi
P_\text{L} \psi\rangle_\text{F} \neq \langle \bar\psi P_\text{R}
\psi\rangle_\text{F}$.  The Green's function is found via $G^{-1} = G_0^{-1} -
\Sigma$ where for example the self-energy of left electron, $\Sigma_{e_L}$, is
\begin{widetext}
\begin{equation}
    \label{eq:37}
    \Sigma_{e_L} = \frac{4 G_F}{\sqrt 2}\left[ 2g_L^e\left( g_L^e\gamma^\mu P_{\text{L}} \langle
        e\bar e\rangle_0 \gamma_\mu P_{\text{L}}  +  \sum_\psi g_{L,R}^\psi \gamma^\mu P_{\text{L}} \langle \bar\psi
        \gamma_\mu P_{\text{L,R}}\psi\rangle_0 \right) + \frac12\gamma^\mu P_{\text{L}} \langle \nu\bar\nu\rangle_0
      \gamma_\mu P_{\text{L}}\right]\equiv \delta \mu_{e_L} \gamma_0 P_{\text{L}}\;.
\end{equation}
\end{widetext}
Expression~(\ref{eq:37}) does not depend on momentum, the thermal averages
$\langle \psi\bar \psi\rangle_0$ and $\langle \bar\psi \gamma_\mu
P_{\text{L,R}}\psi\rangle_0$ are proportional to the particle-antiparticle
asymmetry (see e.g.~\cite{Notzold:87}).  To compute $\delta \mu_{e_L}$ one
should substitute in~(\ref{eq:37}) the thermal averages, summarized in the
Table~1 in Appendix~B.  As a result, for example the electron propagator
becomes
\begin{equation}
  \label{eq:61}
  G = \frac1{\gamma_0\bigl(\delta\mu_{e_{\rm L}}P_{\text{L}} + \delta\mu_{e_{\rm R}}P_{\text{R}}
    + \mu_\text{tree}\bigr) + \slashed{p} + m_e}\;,
\end{equation}
i.e.  the \emph{dispersion relation} of electrons change when taking into
account Fermi corrections (c.f.~\cite{Notzold:87}, \footnote{Technically, this
  is close to the appearance of a chiral shift parameter in the dispersion
  relation of particles in the magnetic fields, discussed in the context of
  compact stars and heavy ion collisions (see~\cite{Gorbar:11b} and references
  therein). However, the physically effect we discuss is of course very
  different.}). Indeed, from $(\slashed{p} - m - \Sigma)\psi = 0$ we see that
the ``on-shell conditions'': $(\omega-\mu_\text{tree})^2= p^2 + m^2$ gets
shifted for left (right) particles by $2(\omega-\mu_\text{tree})
\delta\mu_{L,R}$ (in the limit $\delta\mu_{L,R} \ll \omega$), where
$\mu_\text{tree} = (\lambda_Q - \lambda_e)$.  In the limit $m_e/T\to 0$.
Eq.~(\ref{eq:61}) splits into the sum of free propagators in the
form~(\ref{eq:5}) \emph{with different chemical potentials} $\mu_{L,R}$.

This difference can give rise to a parity-odd term in polarization operator of
photons~\cite{Redlich:1984md}.  Indeed the polarization operator that was
parity-even when computed with respect to the density matrix $\hat\varrho_0$
acquires a parity-odd part when averaged with respect to the $\hat\varrho_F$.
The lowest order weak corrections are represented by two
diagrams,~\ref{fig:vertex} and \ref{fig:bubble}.  The computation of the
diagram~\ref{fig:bubble} is quite similar to that of~\cite{Redlich:1984md} and
gives \emph{a non-zero} $\Pi_2(0) = \frac{\alpha}{2\pi} \sum_f q_f^2 (\delta
\mu_{f_L} - \delta \mu_{f_R})$:
\begin{equation}
  \label{eq:34}
  \Pi_2(0) =  \frac{\alpha}{2\pi} \frac{4G_F}{\sqrt2} \Bigl[c_{L_\alpha} L_\alpha
  + c_B B \Bigr]\;,
\end{equation}
where coefficients $c_{L_\alpha}, c_B\sim \mathcal{O}(1)$ depend on the
fermionic content of the plasma (details are summarized in Appendix)
\footnote{Unlike the effect claimed in~\cite{Semikoz:03} this result does not
  require spatial (temporal) variations of lepton asymmetry and depends only
  on its uniform value.}.  The diagram~\ref{fig:vertex} does not contribute to
the $\Pi_2(0)$ as it can be cut into two diagrams~\ref{fig:Redlich} along the
vertical dotted line, each of which is \emph{at least} first order in momentum
(Eq.~(\ref{eq:63})).

Although the Fermi theory~(\ref{eq:60}) is not renormalizable, the
result~(\ref{eq:34}) is given by the non-divergent part of the
diagram~\ref{fig:self-energy-graph} and is expressed in terms of well-defined
physical quantities (c.f.~\cite{Notzold:87}).

\medskip

\noindent \textbf{Chern-Simons coefficient at two loops and
  ``non-renormalization theorems'' .}  The diagram~\ref{fig:mu} is similar to
the \emph{triangular} diagram, responsible e.g. for $\pi^0\to 2\gamma$ decay
(with $\Delta \mu \bar\psi \gamma_0\gamma_5 \psi$ playing the role of the only
non-zero component of the chiral current, describing
pion)~\cite{Adler:1969gk,Bell:1969ts,Treiman:85}.  It is well known that the
axial anomaly should be calculated at one loop only and is not renormalized by
higher-loop
corrections~\cite{Adler:69,Itoyama:1982up,Contreras:1987ku,Qian:94}, also at
finite temperature and density. At the same time our result becomes non-zero
only at two loops.  There is, however, no contradiction. What is
non-renormalized for the chiral anomaly is the numerical coefficient in front
of the proper combination of external fields, (e.g. $\frac{\alpha}{2\pi}$ in
Eq.~(\ref{eq:4})). In our case this coefficient is also not renormalized.  The
structure of the parity-odd one loop term has the same form at tree-level and
at one-loop in $G_F$: $\Pi_2(0) = \frac{\alpha}{2\pi} (\Delta\mu_\text{tree} +
\delta \mu) A\cdot B$. The numerical coefficient is dictated by the axial
anomaly; $\Delta\mu_\text{tree}$ is a possible difference of chemical
potentials present at tree-level (zero in our case); and $\delta\mu$ is the
\emph{shift} generated by the diagram \ref{fig:bubble}.  The structure of the
$\Pi_2(0)$ term therefore remains the same as in~\cite{Redlich:1984md} with
the \emph{total difference of chemical potential} $(\Delta\mu_\text{tree} +
\delta \mu)$.

Also a four dimensional theory at finite temperatures can be regarded as a
three dimensional Eucledian model albeit with the \emph{infinite number of
  particles} -- each Matsubara mode of a fermion becomes a ``particle'' with
the mass $\omega_n = \pi (2n+1) T$, $n\in\mathbb{Z}$. Therefore (as it was
argued in~\cite{Laine:99}) our result may seem to be in contradiction with the
``Coleman and Hill (CH) theorem''~\cite{Coleman:85} that states that in any
Eucledian three-dimensional gauge theory without massless particles $\Pi_2(0)
= \sum_f \frac{q_f^2}{4\pi}\frac{m_f}{|m_f|}$ and is \emph{exact at one loop}.
However, the presence of the infinite number of modes changes the situation,
as can be seen already in the simplest chiral gauge theory, if one computes
the $\Pi_2(0)$ in the Matsubara formalism (see e.g.~\cite{Tsokos:85}).
Formally, considering the left- and right-chiral particles as fermions with
``complex mass'' $m_n \equiv (\omega_n -i \mu_{L,R})$, and applying directly
to results of~\cite{Coleman:85} one would arrive to an undefined expression $
\Pi_2(0) = \frac{e^2}{4\pi} \sum\limits_{n\in\mathbb{Z}} \frac{\omega_n -
  i\mu_L}{\sqrt{(\omega_n - i\mu_L)^2}} - \frac{\omega_n -
  i\mu_R}{\sqrt{(\omega_n - i\mu_R)^2}}\;.  $ The reason why this happens is
clear: the degree of divergence of the diagrams is different in 3 and 4
dimensions (hence the infinite sum over $n$). In particular, if we
\emph{first} summed over the Matsubara frequencies and \emph{then} integrating
over momentum (or if one uses dimensional regularization in 3-momentum
integral and then takes a limit $d\to 3$,~\cite{Tsokos:85}, see
also~\cite{Niemi:86}) one obtains a well-defined answer
of~\cite{Redlich:1984md}

Moreover, the CH theorem uses the fact that the 3-point vertex
$\Gamma^{(3)}(p_1\dots) =\mathcal{O}(p_1)$. This is not true in our case, as
the diagram~\ref{fig:mu} becomes \emph{linearly divergent} in 4 dimensions and
therefore the shift of integration momentum by any fixed vector $k$ changes
its parity-odd part by a finite amount $\propto \Delta\mu\epsilon^{ijn}A_j k_n
A_j$~\cite{Abers:1971pa,Jackiw:72}.

\medskip

\noindent\textbf{Ground state.}  The presence of $\Pi_2(0)\neq 0$ leads to
the generation of magnetic fields.  The Chern-Simons number $N_\cs\sim k A^2$
will increase until it reaches $\frac{\alpha}{\pi}N_\cs \sim (n_L - n_R) \sim
G_F\Ltot$ (see e.g.~\cite{Rubakov:86}). At fixed $N_\cs$ the magnetic field
tends to increase its wavelength to decrease the total energy ($B^2 \sim k
N_\cs$). As a result, the system does not have a thermodynamic (infinite
volume) limit (c.f.~\cite{Rubakov:86}): the value of the field and the scale
of the inhomogeneity will be determined by the size of the system.  It is
clear, however, that in realistic systems establishing of the long-range field
is a complicated process (see e.g.~\cite{Boyarsky:11a}), greatly affected by
the dissipative processes and by existence of different relaxation channels of
$N_\cs$ (resistivity of plasma, energy radiation, turbulence, etc., see
e.g.~\cite{Banerjee:03,Jedamzik:96,Subramanian:97}).  This may significantly
affect subsequent evolution and even the final state of the system.

\medskip

\noindent \textbf{Discussion.} In this work we demonstrated that the Standard
Model plasma at finite densities of lepton and baryon numbers becomes unstable
and tends to develop large scale magnetic fields.  We considered
electrodynamics plus Fermi theory~(\ref{eq:60}), a description of weak
interactions valid when $e^{-m_W/T} \lesssim (T/m_W)^2$, i.e. at $T\lesssim
40$~GeV.  At higher temperatures one should consider full electroweak theory
and perform two-loop computations of $\Pi_2(0)$. At even higher temperatures
(in the symmetric phase) one should analyze hypermagnetic fields.  We leave
these analyses for future works. We expect however that our conclusion about
the instability of a homogeneous state will hold.

Below we discuss several realistic systems in which the effects discussed here
can become important.  As a first example, let us consider the primordial
plasma at radiation dominated epoch.  Eq.~(\ref{eq:34}) gives $\Pi_2(0) \sim
c\times \alpha (G_FT^3) \eta_{L,B} $ where $\eta_{L,B}<1$ is the ratio of the
total lepton (baryon) number to the number of photons, $n_\gamma=
\frac{6\zeta(3)}{\pi^2} T^3$ and the numerical coefficient $c\approx 2.5\times
10^{-2}$. We see that the CS coefficient decreases with temperature fast and
therefore the effect is the strongest at high temperatures $T\lesssim
m_W$\footnote{Higher-order couplings between fermions and photons, such as
  those leading to anomalous magnetic moment ($\sim G_F m_f \bar\psi
  \sigma_{\mu\nu}\psi F^{\mu\nu}$) are suppressed compared to the considered
  corrections as $\mathcal{O}(m_f/\text{Energy})$. }.

The instability starts to develop at scales $k\sim \Pi_2(0)$ and the magnetic
field initially growth as $e^\beta$ where $\beta \equiv k^2 t/\sigma$ (see
e.g.~\cite{Joyce:97,Boyarsky:11a}). The conductivity of the plasma is $\sigma
\sim \mathcal{O}(10^2)T$~\cite{Baym:1997gq}. The requirement for an
instability to develop over the characteristic time of temperature change
(i.e. Hubble time) is: $\beta(T) \approx
2.0\parfrac{T}{m_W}^3\parfrac{\eta_{L,B}}{10^{-2}}^2 > 1$. We see that the
measured value of \emph{baryon asymmetry} $\eta_B \sim 6.0\times
10^{-10}$~\cite{WMAP7} is too small to trigger any instability. The situation
is different for lepton asymmetry where only the upper bounds at the epoch of
primordial nucleosynthesis exist: $|\eta_L|\lesssim \text{few}\times 10^{-2}$
~\cite{Serpico:05}. At earlier epochs even $\eta_L \sim 1$ is possible (if
this lepton asymmetry disappears later). This is the case e.g. in the $\nu$MSM
(see~\cite{Boyarsky:09a} for review), where the lepton asymmetry keeps being
generated below the sphaleron freeze-out temperature and as a result may reach
the levels $\eta_L \sim 10^{-2} \div 10^{-1}$ before it disappears at
$T\sim{}$~few GeV~\cite{Shaposhnikov:08a}.  We see that significant magnetic
fields can develop in this case, which can play an important role for analysis
of the cosmological implications of the $\nu$MSM.

As a next application we consider a high density degenerate electron plasma
(appearing e.g. in white dwarfs and neutron
stars,~\cite{Shapiro-Teukolsky}). Notice, that our consideration remains valid
in this regime, as Eq.~(\ref{eq:37}) makes no assumption about the relation
between mass, temperature and chemical potential of the particles. Only the
numerical coefficient in Eq.~(\ref{eq:34}) changes and we checked that it is
non-zero and $\mathcal{O}(1)$. The same relation $\Pi_2(0)\sim
\frac{\alpha}{4\pi}G_F \Ltot$ holds, however now $\Ltot = n_e$ (density of
electrons) that can be quite essential, reaching $10^{30}\div
10^{35}~\mathrm{cm}^{-3}$ in the crust of neutron stars~\cite{Chamel:08}. The
corresponding scale of the instability $k\sim \Pi_2(0)$ is then in (sub)km
size and the time of its development is much shorter than the lifetime of the
star.

\medskip

\noindent\textbf{To summarize:} in this work we discussed a previously unknown
effect that occurs in the Standard Model at finite temperature and density. It
implies that a number of processes in the early Universe can be affected,
including cosmological phase transition, baryogenesis, dark matter
production. This effect may in particular lead to the generation of
horizon-scale helical cosmic magnetic fields \emph{purely within the Standard
  Model}. Such fields may survive till present and serve as seeds for the
observed magnetic fields in galaxies and clusters. The effect may also be
important for explanation of physics of compact stars.


\acknowledgements{We acknowledge very helpful conversations with
  J.~Fr\"ohlich, S.~Khlebnikov 
  and M.~Laine. This work was supported in part by the Swiss National Science
  Foundation.}

\bibliography{preamble,astro,cmf,combined_numsm,mybooks,axion,anomalies} %

\let\jnlstyle=\rm\def\jref#1{{\jnlstyle#1}}\def\aj{\jref{AJ}}
  \def\araa{\jref{ARA\&A}} \def\apj{\jref{ApJ}\ } \def\apjl{\jref{ApJ}\ }
  \def\apjs{\jref{ApJS}} \def\ao{\jref{Appl.~Opt.}} \def\apss{\jref{Ap\&SS}}
  \def\aap{\jref{A\&A}} \def\aapr{\jref{A\&A~Rev.}} \def\aaps{\jref{A\&AS}}
  \def\azh{\jref{AZh}} \def\baas{\jref{BAAS}} \def\jrasc{\jref{JRASC}}
  \def\memras{\jref{MmRAS}} \def\mnras{\jref{MNRAS}\ }
  \def\pra{\jref{Phys.~Rev.~A}\ } \def\prb{\jref{Phys.~Rev.~B}\ }
  \def\prc{\jref{Phys.~Rev.~C}\ } \def\prd{\jref{Phys.~Rev.~D}\ }
  \def\pre{\jref{Phys.~Rev.~E}} \def\prl{\jref{Phys.~Rev.~Lett.}}
  \def\pasp{\jref{PASP}} \def\pasj{\jref{PASJ}} \def\qjras{\jref{QJRAS}}
  \def\skytel{\jref{S\&T}} \def\solphys{\jref{Sol.~Phys.}}
  \def\sovast{\jref{Soviet~Ast.}} \def\ssr{\jref{Space~Sci.~Rev.}}
  \def\zap{\jref{ZAp}} \def\nat{\jref{Nature}\ } \def\iaucirc{\jref{IAU~Circ.}}
  \def\aplett{\jref{Astrophys.~Lett.}}
  \def\apspr{\jref{Astrophys.~Space~Phys.~Res.}}
  \def\bain{\jref{Bull.~Astron.~Inst.~Netherlands}}
  \def\fcp{\jref{Fund.~Cosmic~Phys.}} \def\gca{\jref{Geochim.~Cosmochim.~Acta}}
  \def\grl{\jref{Geophys.~Res.~Lett.}} \def\jcp{\jref{J.~Chem.~Phys.}}
  \def\jgr{\jref{J.~Geophys.~Res.}}
  \def\jqsrt{\jref{J.~Quant.~Spec.~Radiat.~Transf.}}
  \def\memsai{\jref{Mem.~Soc.~Astron.~Italiana}}
  \def\nphysa{\jref{Nucl.~Phys.~A}} \def\physrep{\jref{Phys.~Rep.}}
  \def\physscr{\jref{Phys.~Scr}} \def\planss{\jref{Planet.~Space~Sci.}}
  \def\procspie{\jref{Proc.~SPIE}} \let\astap=\aap \let\apjlett=\apjl
  \let\apjsupp=\apjs \let\applopt=\ao
\begin{thebibliography}{43}
\expandafter\ifx\csname natexlab\endcsname\relax\def\natexlab#1{#1}\fi
\expandafter\ifx\csname bibnamefont\endcsname\relax
  \def\bibnamefont#1{#1}\fi
\expandafter\ifx\csname bibfnamefont\endcsname\relax
  \def\bibfnamefont#1{#1}\fi
\expandafter\ifx\csname citenamefont\endcsname\relax
  \def\citenamefont#1{#1}\fi
\expandafter\ifx\csname url\endcsname\relax
  \def\url#1{\texttt{#1}}\fi
\expandafter\ifx\csname urlprefix\endcsname\relax\def\urlprefix{URL }\fi
\providecommand{\bibinfo}[2]{#2}
\providecommand{\eprint}[2][]{\url{#2}}

\bibitem[{\citenamefont{Furry}(1937)}]{Furry:1937zz}
\bibinfo{author}{\bibfnamefont{W.}~\bibnamefont{Furry}},
  \bibinfo{journal}{Phys.Rev.} \textbf{\bibinfo{volume}{51}},
  \bibinfo{pages}{125} (\bibinfo{year}{1937}).

\bibitem[{\citenamefont{Linde}(1980)}]{Linde:1980ts}
\bibinfo{author}{\bibfnamefont{A.~D.} \bibnamefont{Linde}},
  \bibinfo{journal}{Phys.Lett.} \textbf{\bibinfo{volume}{B96}},
  \bibinfo{pages}{289} (\bibinfo{year}{1980}).

\bibitem[{\citenamefont{Fradkin}(1965)}]{Fradkin:65}
\bibinfo{author}{\bibfnamefont{E.}~\bibnamefont{Fradkin}}, in
  \emph{\bibinfo{booktitle}{Proc. PN Lebedev Phys. Inst}}
  (\bibinfo{year}{1965}), vol.~\bibinfo{volume}{29}.

\bibitem[{\citenamefont{Redlich and Wijewardhana}(1985)}]{Redlich:1984md}
\bibinfo{author}{\bibfnamefont{A.~N.} \bibnamefont{Redlich}} \bibnamefont{and}
  \bibinfo{author}{\bibfnamefont{L.~C.~R.} \bibnamefont{Wijewardhana}},
  \bibinfo{journal}{Phys. Rev. Lett.} \textbf{\bibinfo{volume}{54}},
  \bibinfo{pages}{970} (\bibinfo{year}{1985}).

\bibitem[{\citenamefont{Rubakov and Tavkhelidze}(1985)}]{Rubakov:86b}
\bibinfo{author}{\bibfnamefont{V.~A.} \bibnamefont{Rubakov}} \bibnamefont{and}
  \bibinfo{author}{\bibfnamefont{A.~N.} \bibnamefont{Tavkhelidze}},
  \bibinfo{journal}{Phys. Lett.} \textbf{\bibinfo{volume}{B165}},
  \bibinfo{pages}{109} (\bibinfo{year}{1985}).

\bibitem[{\citenamefont{Rubakov}(1986)}]{Rubakov:86}
\bibinfo{author}{\bibfnamefont{V.~A.} \bibnamefont{Rubakov}},
  \bibinfo{journal}{Prog. Theor. Phys.} \textbf{\bibinfo{volume}{75}},
  \bibinfo{pages}{366} (\bibinfo{year}{1986}).

\bibitem[{\citenamefont{Deryagin et~al.}(1986)\citenamefont{Deryagin,
  Grigoriev, and Rubakov}}]{Deryagin:1986kx}
\bibinfo{author}{\bibfnamefont{D.}~\bibnamefont{Deryagin}},
  \bibinfo{author}{\bibfnamefont{D.~Y.} \bibnamefont{Grigoriev}},
  \bibnamefont{and} \bibinfo{author}{\bibfnamefont{V.}~\bibnamefont{Rubakov}},
  \bibinfo{journal}{Phys.Lett.} \textbf{\bibinfo{volume}{B178}},
  \bibinfo{pages}{385} (\bibinfo{year}{1986}).

\bibitem[{\citenamefont{Joyce and Shaposhnikov}(1997)}]{Joyce:97}
\bibinfo{author}{\bibfnamefont{M.}~\bibnamefont{Joyce}} \bibnamefont{and}
  \bibinfo{author}{\bibfnamefont{M.~E.} \bibnamefont{Shaposhnikov}},
  \bibinfo{journal}{Phys. Rev. Lett.} \textbf{\bibinfo{volume}{79}},
  \bibinfo{pages}{1193} (\bibinfo{year}{1997}), \eprint{astro-ph/9703005}.

\bibitem[{\citenamefont{Tsokos}(1985)}]{Tsokos:85}
\bibinfo{author}{\bibfnamefont{K.}~\bibnamefont{Tsokos}},
  \bibinfo{journal}{Phys.Lett.} \textbf{\bibinfo{volume}{B157}},
  \bibinfo{pages}{413} (\bibinfo{year}{1985}).

\bibitem[{\citenamefont{Niemi and Semenoff}(1985)}]{Niemi:85}
\bibinfo{author}{\bibfnamefont{A.}~\bibnamefont{Niemi}} \bibnamefont{and}
  \bibinfo{author}{\bibfnamefont{G.}~\bibnamefont{Semenoff}},
  \bibinfo{journal}{Phys.Rev.Lett.} \textbf{\bibinfo{volume}{54}},
  \bibinfo{pages}{2166} (\bibinfo{year}{1985}).

\bibitem[{\citenamefont{Niemi}(1986)}]{Niemi:86}
\bibinfo{author}{\bibfnamefont{A.}~\bibnamefont{Niemi}},
  \bibinfo{journal}{Phys.Rev.Lett.} \textbf{\bibinfo{volume}{57}},
  \bibinfo{pages}{1102} (\bibinfo{year}{1986}).

\bibitem[{\citenamefont{Vilenkin}(1980)}]{Vilenkin:80a}
\bibinfo{author}{\bibfnamefont{A.}~\bibnamefont{Vilenkin}},
  \bibinfo{journal}{Phys. Rev.} \textbf{\bibinfo{volume}{D22}},
  \bibinfo{pages}{3080} (\bibinfo{year}{1980}).

\bibitem[{\citenamefont{Alekseev et~al.}(1998)\citenamefont{Alekseev, Cheianov,
  and Frohlich}}]{Alekseev:98a}
\bibinfo{author}{\bibfnamefont{A.~Y.} \bibnamefont{Alekseev}},
  \bibinfo{author}{\bibfnamefont{V.~V.} \bibnamefont{Cheianov}},
  \bibnamefont{and} \bibinfo{author}{\bibfnamefont{J.}~\bibnamefont{Frohlich}},
  \bibinfo{journal}{Phys. Rev. Lett.} \textbf{\bibinfo{volume}{81}},
  \bibinfo{pages}{3503} (\bibinfo{year}{1998}), \eprint{cond-mat/9803346}.

\bibitem[{\citenamefont{Fr\"{o}hlich and Pedrini}(2000)}]{Frohlich:2000en}
\bibinfo{author}{\bibfnamefont{J.}~\bibnamefont{Fr\"{o}hlich}}
  \bibnamefont{and} \bibinfo{author}{\bibfnamefont{B.}~\bibnamefont{Pedrini}},
  in \emph{\bibinfo{booktitle}{Mathematical Physics 2000}}, edited by
  \bibinfo{editor}{\bibfnamefont{A.~S.} \bibnamefont{Fokas}},
  \bibinfo{editor}{\bibfnamefont{A.}~\bibnamefont{Grigoryan}},
  \bibinfo{editor}{\bibfnamefont{T.}~\bibnamefont{Kibble}}, \bibnamefont{and}
  \bibinfo{editor}{\bibfnamefont{B.}~\bibnamefont{Zegarlinski}}
  (\bibinfo{publisher}{World Scientific Publishing Company},
  \bibinfo{year}{2000}), International Conference on Mathematical Physics 2000,
  Imperial college (London), \eprint{hep-th/0002195}.

\bibitem[{\citenamefont{Adler}(1969)}]{Adler:1969gk}
\bibinfo{author}{\bibfnamefont{S.~L.} \bibnamefont{Adler}},
  \bibinfo{journal}{Phys. Rev.} \textbf{\bibinfo{volume}{177}},
  \bibinfo{pages}{2426} (\bibinfo{year}{1969}).

\bibitem[{\citenamefont{Bell and Jackiw}(1969)}]{Bell:1969ts}
\bibinfo{author}{\bibfnamefont{J.~S.} \bibnamefont{Bell}} \bibnamefont{and}
  \bibinfo{author}{\bibfnamefont{R.}~\bibnamefont{Jackiw}},
  \bibinfo{journal}{Nuovo Cim.} \textbf{\bibinfo{volume}{A60}},
  \bibinfo{pages}{47} (\bibinfo{year}{1969}).

\bibitem[{\citenamefont{Treiman et~al.}(1985)\citenamefont{Treiman, Jackiw,
  Witten, and Zumino}}]{Treiman:85}
\bibinfo{author}{\bibfnamefont{S.}~\bibnamefont{Treiman}},
  \bibinfo{author}{\bibfnamefont{R.}~\bibnamefont{Jackiw}},
  \bibinfo{author}{\bibfnamefont{E.}~\bibnamefont{Witten}}, \bibnamefont{and}
  \bibinfo{author}{\bibfnamefont{B.}~\bibnamefont{Zumino}},
  \emph{\bibinfo{title}{Current algebra and anomalies}}, Princeton series in
  physics (\bibinfo{publisher}{Princeton University Press},
  \bibinfo{year}{1985}), ISBN \bibinfo{isbn}{9780691083988}.

\bibitem[{\citenamefont{Gynther}(2003)}]{Gynther:03}
\bibinfo{author}{\bibfnamefont{A.}~\bibnamefont{Gynther}},
  \bibinfo{journal}{Phys.Rev.} \textbf{\bibinfo{volume}{D68}},
  \bibinfo{pages}{016001} (\bibinfo{year}{2003}), \eprint{hep-ph/0303019}.

\bibitem[{\citenamefont{Kuzmin et~al.}(1985)\citenamefont{Kuzmin, Rubakov, and
  Shaposhnikov}}]{Kuzmin:85}
\bibinfo{author}{\bibfnamefont{V.~A.} \bibnamefont{Kuzmin}},
  \bibinfo{author}{\bibfnamefont{V.~A.} \bibnamefont{Rubakov}},
  \bibnamefont{and} \bibinfo{author}{\bibfnamefont{M.~E.}
  \bibnamefont{Shaposhnikov}}, \bibinfo{journal}{Phys. Lett.}
  \textbf{\bibinfo{volume}{B155}}, \bibinfo{pages}{36} (\bibinfo{year}{1985}).

\bibitem[{\citenamefont{Boyarsky et~al.}(2012)\citenamefont{Boyarsky,
  Fr{\"{o}}lich, and Ruchayskiy}}]{Boyarsky:11a}
\bibinfo{author}{\bibfnamefont{A.}~\bibnamefont{Boyarsky}},
  \bibinfo{author}{\bibfnamefont{J.}~\bibnamefont{Fr{\"{o}}lich}},
  \bibnamefont{and}
  \bibinfo{author}{\bibfnamefont{O.}~\bibnamefont{Ruchayskiy}},
  \bibinfo{journal}{Phys.Rev.Lett.} \textbf{\bibinfo{volume}{108}},
  \bibinfo{pages}{031301} (\bibinfo{year}{2012}), \eprint{1109.3350}.

\bibitem[{\citenamefont{Notzold and Raffelt}(1988)}]{Notzold:87}
\bibinfo{author}{\bibfnamefont{D.}~\bibnamefont{Notzold}} \bibnamefont{and}
  \bibinfo{author}{\bibfnamefont{G.}~\bibnamefont{Raffelt}},
  \bibinfo{journal}{Nucl. Phys.} \textbf{\bibinfo{volume}{B307}},
  \bibinfo{pages}{924} (\bibinfo{year}{1988}).

\bibitem[{\citenamefont{Adler and Bardeen}(1969)}]{Adler:69}
\bibinfo{author}{\bibfnamefont{S.~L.} \bibnamefont{Adler}} \bibnamefont{and}
  \bibinfo{author}{\bibfnamefont{W.~A.} \bibnamefont{Bardeen}},
  \bibinfo{journal}{Phys.Rev.} \textbf{\bibinfo{volume}{182}},
  \bibinfo{pages}{1517} (\bibinfo{year}{1969}).

\bibitem[{\citenamefont{Itoyama and Mueller}(1983)}]{Itoyama:1982up}
\bibinfo{author}{\bibfnamefont{H.}~\bibnamefont{Itoyama}} \bibnamefont{and}
  \bibinfo{author}{\bibfnamefont{A.~H.} \bibnamefont{Mueller}},
  \bibinfo{journal}{Nucl.Phys.} \textbf{\bibinfo{volume}{B218}},
  \bibinfo{pages}{349} (\bibinfo{year}{1983}).

\bibitem[{\citenamefont{Contreras and Loewe}(1988)}]{Contreras:1987ku}
\bibinfo{author}{\bibfnamefont{C.}~\bibnamefont{Contreras}} \bibnamefont{and}
  \bibinfo{author}{\bibfnamefont{M.}~\bibnamefont{Loewe}},
  \bibinfo{journal}{Z.Phys.} \textbf{\bibinfo{volume}{C40}},
  \bibinfo{pages}{253} (\bibinfo{year}{1988}).

\bibitem[{\citenamefont{Qian et~al.}(1994)\citenamefont{Qian, Su, and
  Yu}}]{Qian:94}
\bibinfo{author}{\bibfnamefont{Z.-X.} \bibnamefont{Qian}},
  \bibinfo{author}{\bibfnamefont{R.-K.} \bibnamefont{Su}}, \bibnamefont{and}
  \bibinfo{author}{\bibfnamefont{P.}~\bibnamefont{Yu}},
  \bibinfo{journal}{Z.Phys.} \textbf{\bibinfo{volume}{C63}},
  \bibinfo{pages}{651} (\bibinfo{year}{1994}).

\bibitem[{\citenamefont{Laine and Shaposhnikov}(1999)}]{Laine:99}
\bibinfo{author}{\bibfnamefont{M.}~\bibnamefont{Laine}} \bibnamefont{and}
  \bibinfo{author}{\bibfnamefont{M.~E.} \bibnamefont{Shaposhnikov}},
  \bibinfo{journal}{Phys.Lett.} \textbf{\bibinfo{volume}{B463}},
  \bibinfo{pages}{280} (\bibinfo{year}{1999}), \eprint{hep-th/9907194}.

\bibitem[{\citenamefont{Coleman and Hill}(1985)}]{Coleman:85}
\bibinfo{author}{\bibfnamefont{S.~R.} \bibnamefont{Coleman}} \bibnamefont{and}
  \bibinfo{author}{\bibfnamefont{B.~R.} \bibnamefont{Hill}},
  \bibinfo{journal}{Phys.Lett.} \textbf{\bibinfo{volume}{B159}},
  \bibinfo{pages}{184} (\bibinfo{year}{1985}).

\bibitem[{\citenamefont{Abers et~al.}(1971)\citenamefont{Abers, Dicus, and
  Teplitz}}]{Abers:1971pa}
\bibinfo{author}{\bibfnamefont{E.}~\bibnamefont{Abers}},
  \bibinfo{author}{\bibfnamefont{D.}~\bibnamefont{Dicus}}, \bibnamefont{and}
  \bibinfo{author}{\bibfnamefont{V.}~\bibnamefont{Teplitz}},
  \bibinfo{journal}{Phys.Rev.} \textbf{\bibinfo{volume}{D3}},
  \bibinfo{pages}{485} (\bibinfo{year}{1971}).

\bibitem[{\citenamefont{Jackiw}(1972)}]{Jackiw:72}
\bibinfo{author}{\bibfnamefont{R.}~\bibnamefont{Jackiw}},
  \emph{\bibinfo{title}{Field theoretic investigations in current algebra}},
  \bibinfo{howpublished}{in~\cite{Treiman:85}} (\bibinfo{year}{1972}).

\bibitem[{\citenamefont{Banerjee and Jedamzik}(2003)}]{Banerjee:03}
\bibinfo{author}{\bibfnamefont{R.}~\bibnamefont{Banerjee}} \bibnamefont{and}
  \bibinfo{author}{\bibfnamefont{K.}~\bibnamefont{Jedamzik}},
  \bibinfo{journal}{Phys. Rev. Lett.} \textbf{\bibinfo{volume}{91}},
  \bibinfo{pages}{251301} (\bibinfo{year}{2003}), \eprint{astro-ph/0306211}.

\bibitem[{\citenamefont{Jedamzik et~al.}(1998)\citenamefont{Jedamzik,
  Katalinic, and Olinto}}]{Jedamzik:96}
\bibinfo{author}{\bibfnamefont{K.}~\bibnamefont{Jedamzik}},
  \bibinfo{author}{\bibfnamefont{V.}~\bibnamefont{Katalinic}},
  \bibnamefont{and} \bibinfo{author}{\bibfnamefont{A.~V.}
  \bibnamefont{Olinto}}, \bibinfo{journal}{Phys. Rev.}
  \textbf{\bibinfo{volume}{D57}}, \bibinfo{pages}{3264} (\bibinfo{year}{1998}),
  \eprint{astro-ph/9606080}.

\bibitem[{\citenamefont{Subramanian and Barrow}(1998)}]{Subramanian:97}
\bibinfo{author}{\bibfnamefont{K.}~\bibnamefont{Subramanian}} \bibnamefont{and}
  \bibinfo{author}{\bibfnamefont{J.~D.} \bibnamefont{Barrow}},
  \bibinfo{journal}{Phys.Rev.} \textbf{\bibinfo{volume}{D58}},
  \bibinfo{pages}{083502} (\bibinfo{year}{1998}), \eprint{astro-ph/9712083}.

\bibitem[{\citenamefont{Baym and Heiselberg}(1997)}]{Baym:1997gq}
\bibinfo{author}{\bibfnamefont{G.}~\bibnamefont{Baym}} \bibnamefont{and}
  \bibinfo{author}{\bibfnamefont{H.}~\bibnamefont{Heiselberg}},
  \bibinfo{journal}{Phys. Rev.} \textbf{\bibinfo{volume}{D56}},
  \bibinfo{pages}{5254} (\bibinfo{year}{1997}), \eprint{astro-ph/9704214}.

\bibitem[{\citenamefont{{Komatsu} et~al.}(2011)}]{WMAP7}
\bibinfo{author}{\bibfnamefont{E.}~\bibnamefont{{Komatsu}}}
  \bibnamefont{et~al.} (\bibinfo{collaboration}{WMAP}),
  \bibinfo{journal}{\apjs} \textbf{\bibinfo{volume}{192}}, \bibinfo{pages}{18}
  (\bibinfo{year}{2011}), \eprint{1001.4538}.

\bibitem[{\citenamefont{Serpico and Raffelt}(2005)}]{Serpico:05}
\bibinfo{author}{\bibfnamefont{P.~D.} \bibnamefont{Serpico}} \bibnamefont{and}
  \bibinfo{author}{\bibfnamefont{G.~G.} \bibnamefont{Raffelt}},
  \bibinfo{journal}{Phys. Rev.} \textbf{\bibinfo{volume}{D71}},
  \bibinfo{pages}{127301} (\bibinfo{year}{2005}), \eprint{astro-ph/0506162}.

\bibitem[{\citenamefont{Boyarsky et~al.}(2009)\citenamefont{Boyarsky,
  Ruchayskiy, and Shaposhnikov}}]{Boyarsky:09a}
\bibinfo{author}{\bibfnamefont{A.}~\bibnamefont{Boyarsky}},
  \bibinfo{author}{\bibfnamefont{O.}~\bibnamefont{Ruchayskiy}},
  \bibnamefont{and}
  \bibinfo{author}{\bibfnamefont{M.}~\bibnamefont{Shaposhnikov}},
  \bibinfo{journal}{Ann. Rev. Nucl. Part. Sci.} \textbf{\bibinfo{volume}{59}},
  \bibinfo{pages}{191} (\bibinfo{year}{2009}), \eprint{0901.0011}.

\bibitem[{\citenamefont{{Shaposhnikov}}(2008)}]{Shaposhnikov:08a}
\bibinfo{author}{\bibfnamefont{M.}~\bibnamefont{{Shaposhnikov}}},
  \bibinfo{journal}{JHEP} \textbf{\bibinfo{volume}{08}}, \bibinfo{pages}{008}
  (\bibinfo{year}{2008}), \eprint{0804.4542}.

\bibitem[{\citenamefont{Shapiro and Teukolsky}(2008)}]{Shapiro-Teukolsky}
\bibinfo{author}{\bibfnamefont{S.}~\bibnamefont{Shapiro}} \bibnamefont{and}
  \bibinfo{author}{\bibfnamefont{S.}~\bibnamefont{Teukolsky}},
  \emph{\bibinfo{title}{Black Holes, White Dwarfs and Neutron Stars: The
  Physics of Compact Objects}} (\bibinfo{publisher}{John Wiley \& Sons},
  \bibinfo{year}{2008}), ISBN \bibinfo{isbn}{9783527617678}.

\bibitem[{\citenamefont{Chamel and Haensel}(2008)}]{Chamel:08}
\bibinfo{author}{\bibfnamefont{N.}~\bibnamefont{Chamel}} \bibnamefont{and}
  \bibinfo{author}{\bibfnamefont{P.}~\bibnamefont{Haensel}},
  \bibinfo{journal}{Living Rev. Rel.} \textbf{\bibinfo{volume}{11}},
  \bibinfo{pages}{10} (\bibinfo{year}{2008}), \eprint{0812.3955}.

\bibitem[{\citenamefont{Peskin and Schroeder}(1995)}]{Peskin-Schroeder}
\bibinfo{author}{\bibfnamefont{M.~E.} \bibnamefont{Peskin}} \bibnamefont{and}
  \bibinfo{author}{\bibfnamefont{D.~V.} \bibnamefont{Schroeder}},
  \emph{\bibinfo{title}{An Introduction To Quantum Field Theory (Frontiers in
  Physics)}} (\bibinfo{publisher}{Westview Press}, \bibinfo{year}{1995}), ISBN
  \bibinfo{isbn}{0201503972}.

\bibitem[{\citenamefont{Gorbar et~al.}(2012)\citenamefont{Gorbar, Miransky, and
  Shovkovy}}]{Gorbar:11b}
\bibinfo{author}{\bibfnamefont{E.}~\bibnamefont{Gorbar}},
  \bibinfo{author}{\bibfnamefont{V.}~\bibnamefont{Miransky}}, \bibnamefont{and}
  \bibinfo{author}{\bibfnamefont{I.}~\bibnamefont{Shovkovy}},
  \bibinfo{journal}{Prog.Part.Nucl.Phys.} \textbf{\bibinfo{volume}{67}},
  \bibinfo{pages}{547} (\bibinfo{year}{2012}), \eprint{1111.3401}.

\bibitem[{\citenamefont{Semikoz and Sokoloff}(2004)}]{Semikoz:03}
\bibinfo{author}{\bibfnamefont{V.}~\bibnamefont{Semikoz}} \bibnamefont{and}
  \bibinfo{author}{\bibfnamefont{D.}~\bibnamefont{Sokoloff}},
  \bibinfo{journal}{Phys.Rev.Lett.} \textbf{\bibinfo{volume}{92}},
  \bibinfo{pages}{131301} (\bibinfo{year}{2004}), \eprint{astro-ph/0312567}.

\bibitem[{\citenamefont{Khlebnikov and Shaposhnikov}(1988)}]{Khlebnikov:88}
\bibinfo{author}{\bibfnamefont{S.~Y.} \bibnamefont{Khlebnikov}}
  \bibnamefont{and} \bibinfo{author}{\bibfnamefont{M.~E.}
  \bibnamefont{Shaposhnikov}}, \bibinfo{journal}{Nucl. Phys.}
  \textbf{\bibinfo{volume}{B308}}, \bibinfo{pages}{885} (\bibinfo{year}{1988}).

\end{thebibliography}

\newpage
\appendix

\onecolumngrid

\section{Fermi theory}

For completeness we summarize in this Appendix the definitions of charged and
neutral currents in the Fermi theory (see e.g. Chapter~20
in~\cite{Peskin-Schroeder}).  The \emph{charged currents} are defined as
\begin{equation}
J^{\cc,+}_\mu =\frac1{\sqrt 2}\left( \bar e_L \gamma_\mu \nu_L +
  \bar d_L \gamma_\mu u_L\right) + \mbox{other generations} \;,
\label{eq:65}
\end{equation}
and
\emph{neutral currents} are 
\begin{equation}
 J^\nc_\mu = \sum_{\psi}\bar\psi \gamma^\mu\Bigl
(g_L^\psi P_{\text{L}} + g_R^\psi P_{\text{R}}\Bigr)\psi \;.
\label{eq:74}
\end{equation}
Here $P_{\text{L,R}} = \frac12(1\pm \gamma_5)$ are chiral projectors, charges
$g_L^\psi = (T_3 - \sw Q)$, $g_R^\psi = (-\sw Q)$, $T_3 =\pm \frac12$ is the
3rd generator of $SU(2)$, $Q$ is the \emph{electric charge} and the sum
in~(\ref{eq:74}) goes over fermions in all flavours.

\section{General expression for the Chern-Simons coefficient $\Pi_2(0)$}
\label{sec:general expression}

\begin{table}[!t]
  \centering
  \begin{tabular}{ll}
    \hline
    Charged leptons: & $\langle \bar\ell \gamma^\mu P \ell\rangle =
    \delta^{\mu0} \frac{1}{2}\Delta n_\ell$\\
    & $\langle \ell \bar\ell \rangle = \frac14 \gamma_0 (-\Delta n_\ell)$\\
    \hline
    Neutrinos: & $\langle \bar\nu \gamma^\mu P \nu\rangle =
    \delta^{\mu0} \Delta n_\nu$\\
    &$\langle \nu \bar\nu\rangle = \frac12 P_{\text{L}} \gamma_0 (-\Delta n_\nu)$\\
    \hline
    Quarks: & $\langle \bar q \gamma^\mu P q\rangle =
    \delta^{\mu0} \frac{1}{2}\Delta n_q$\\
    & $\langle q \bar q \rangle = \frac1{12} \gamma_0 (-\Delta n_q)$\\
    \hline
  \end{tabular}
  \caption{Thermal averages of fermions. Particle-antiparticle asymmetry
    is defined as $\Delta n = \frac{g}2\int\frac{d^3\vec p}{(2\pi)^3}
    \left[ f_p - \bar f_p \right]$, $f_p(\bar f_p)$ is the Fermi-Dirac distribution for
    particles (anti-particles) and  the number of
    internal degrees of freedom, $g = 2$ for neutrinos $\nu$; $g=4$ for charged
    leptons $\ell$;
    $g=12$ for quarks $q$.}
  \label{tab:summary}
\end{table}

A general expression for $\Pi_2(0)$ is given through the asymmetries of all
fermions (if some fermions are absent in the plasma, their asymmetry should be
put to zero). 
\begin{equation}
  \label{eq:6}
  \begin{aligned}
    \Pi_2(0) = &\frac\alpha{2\pi}\frac{4G_F}{\sqrt 2}
    \biggl(-\frac29\sum_\alpha \Delta n_{\nu_\alpha } -\frac{31}{36}
    \bigl(1-2\cos(2\theta_W)\bigr) \sum_\alpha \Delta
    n_{\ell_\alpha }\\
    &+\frac1{81}\bigl(17 - 62 \cos(2\theta_W)\bigr) (\Delta n_u + \Delta n_c)\\
    &+\frac{1}{324}\bigl(91 + 134 \cos(2\theta_W)\bigr) (\Delta n_d + \Delta
    n_s)\\
    &+\frac{67}{324}\bigl(1 + 2 \cos(2\theta_W)\bigr) \Delta n_b\biggr)\;,
  \end{aligned}
\end{equation}
(here $\theta_W$ is the Weinberg's angle).

Once the asymmetries of all particles $\Delta n$ are expressed through the
conserved charges $B$ and $L_\alpha$ under the condition of electric
neutrality of the plasma~\cite{Khlebnikov:88}, the expression~(\ref{eq:6})
reduces to the form~(\ref{eq:34}):
\begin{equation}
\label{eq:11}
  \Pi_2(0) =  \frac{\alpha}{2\pi} \frac{4G_F}{\sqrt2} \Bigl[c_{L_\alpha} L_\alpha
  + c_B B \Bigr]\;,
\end{equation}
where the values of coefficients depend on the fermionic content of the
plasma.  For example, if plasma contains 5 quarks (except for the top quark)
and all leptons, then
\begin{equation}
  \label{eq:64}
  c_{L_\alpha} = \frac{8\Bigl(22\cos(2\theta_W) - 45\Bigr)}{621}\quad,\quad c_B = \frac{\Bigl(53\cos(2\theta_W) +430\Bigr)}{621} 
\end{equation}

\end{document}